\documentclass[11pt,a4paper]{article}
\usepackage{jheppub}
\usepackage{lmodern}

\usepackage[T1]{fontenc}
\usepackage{units}
\usepackage{pmboxdraw}
\usepackage{amstext}
\usepackage{amsmath}
\usepackage{amssymb}
\usepackage{graphicx}
\usepackage{epstopdf}
\usepackage{appendix}
\usepackage{setspace}
\usepackage{esint}

\makeatletter
\@ifundefined{showcaptionsetup}{}{%
\PassOptionsToPackage{caption=false}{subfig}}
\usepackage{subfig}
\makeatother
\preprint{FTUV-13-11-29, ~IFIC-13-93}
\title{Improved $\tau$-weapons for Higgs hunting}
\author{G.~Barenboim}\emailAdd{Gabriela.Barenboim@uv.es}
\author{C.~Bosch}\emailAdd{Cristian.Bosch@uv.es}
\author{M.L.~L\'opez-Ib\'a\~nez}\emailAdd{M.Luisa.Lopez-Ibanez@uv.es}
\author{O. Vives}\emailAdd{Oscar.Vives@uv.es}
\affiliation{Departament de F\'{\i}sica Te\`orica and IFIC, Universitat de 
Val\`encia-CSIC, E-46100, Burjassot, Spain.}
\abstract{In this work, we use the results from Higgs searches in the $\gamma\gamma$ and $\tau\tau$ decay channels at LHC and indirect bounds as BR$(B \to X_s \gamma)$  to constrain the parameter space of a generic MSSM Higgs sector. In particular, we include the latest CMS results that look for additional Higgs states with masses up to 1 TeV. We show that the $\tau \tau$ channel is the best and most accurate weapon in the hunt for new Higgs states beyond the Standard Model. We obtain that present experimental results rule out additional neutral Higgs bosons in a generic MSSM below 300~GeV for any value of $\tan \beta$ and, for instance, values of $\tan \beta$ above 30 are only possible for Higgs masses above 600~GeV. ATLAS stored data have the potential to render this bound obsolete in the near future.}
\begin{document}
\maketitle
\section{Introduction}
The main purpose of the LHC, {\it i.e.} to find the Higgs boson and complete the
Standard Model (SM) construction, was recently fulfilled with the
discovery, at ATLAS and CMS, of a bosonic resonance with a mass
$\sim126$~GeV \cite{Aad:2012tfa,Chatrchyan:2012ufa}. The relevance of this discovery can not be understimated because of the key role the Higgs boson plays in the structure of the SM, 
as it provides the
mechanism of electroweak symmetry breaking and generates the masses
for gauge bosons and fermions. Likewise,  in
all the SM extensions other scalar bosons associated to the breaking of the 
electroweak symmetry are present and play an equaly important role.

However, to confirm that this resonance corresponds indeed to the SM Higgs
boson or it belongs to one of the SM extensions, it is necessary to measure in LHC experiments its properties
and couplings with high precision
\cite{Aad:2013wqa,CMS-PAS-HIG-13-005,ATLAS-CONF-2013-034,ATLAS-CONF-2013-030,ATLAS-CONF-2013-013,ATLAS-CONF-2013-012,CMS-PAS-HIG-13-016,CMS-PAS-HIG-13-020,CMS-PAS-HIG-13-022,CMS-PAS-HIG-13-019,CMS-PAS-HIG-12-024}.
At present, the observed production cross section and decay channels
seem to be consistent, within errors, with a Higgs boson in the SM
framework.  But the current experimental precision leaves  the
possibility of this resonance being a Higgs boson of one of the
different extensions of the SM open \cite{Aad:2013wqa,CMS-PAS-HIG-13-005}.
To clarify this issue, further experimental studies on the resonance
properties are needed, together with complementary studies looking for 
new scalar particles
that are usually present in these extensions of the SM.

The prototype SM extension in the Higgs sector is the so-called two
Higgs doublet model (2HdM). In a 2HdM, the Higgs sector is expanded
with the inclusion of a second scalar doublet of opposite
hypercharge. This enlargement of the scalar sector leads to an
increase in the number of physical Higgs states in the spectrum, that
is then composed by two scalar states, one pseudoscalar state and a
charged Higgs boson. In particular, it is well-known that the Higgs
sector of the MSSM is a type II 2HdM \cite{Nilles:1983ge,Haber:1984rc,Djouadi:2005gj}. The type II qualifier refers to
the fact that, at tree level, only one of the doublets couples to
down-type fermions while the second one couples to up-type
fermions. This is one of the classical mechanisms to avoid the
appearance of flavour changing neutral currents (FCNC) at
tree-level. The requirement of holomorficity of the superpotential
together with gauge invariance forces the Higgs sector in a
supersymmetric model to be precisely a type II 2HdM, and this is
the scenario where we will perform our analysis.  The MSSM is the
minimal supersymmetric extension of the Standard Model with respect to
particle content. In particular the Higgs sector of the MSSM is a
type II 2HdM and it is CP-conserving at tree-level \cite{Nilles:1983ge,Haber:1984rc,Djouadi:2005gj}. However, loop
effects involving the complex parameters in the MSSM Lagrangian
violate the tree-level CP-invariance of the MSSM Higgs potential
modifying the tree-level masses, couplings, production rates and decay
widths of Higgs bosons
\cite{Pilaftsis:1999qt,Carena:2000yi,Choi:2000wz,Carena:2001fw,Choi:2001pg,Choi:2002zp}.
In this way, the physical Higgs eigenstates become admixtures of
CP-even and odd states and its couplings to SM particles are modified
accordingly.

In a recent paper \cite{Barenboim:2013bla}, we carried out an analysis in a generic MSSM under the assumption that the observed Higgs state corresponded to the second-lightest Higgs. We were able to eliminate this possibility analytically, using only the diphoton signal strength, $\tau\tau$ production through Higgs and BR($B \to X_s \gamma$). In this work, we follow a similar strategy to scrutinize the allowed areas of parameter space in a complex MSSM in the case the Higgs measured at LHC is the lightest MSSM Higgs boson. We look for the best observables to identify the nature of the Higgs sector and, specially, where it is more appropriate to search for additional Higgs states. 

As our analysis concentrates mainly on the Higgs sector of the MSSM and this sector is affected only by a handful of MSSM parameters, it is possible to perform a general phenomenological analysis in terms of these parameters encompassing all the different MSSM setups. In this context, we fix $m_{H_1}\simeq 126~\mbox{GeV}\leq m_{H_2}, m_{H_3}, m_{H^\pm}$ and use the experimental results to look for acceptable values for these Higgs masses and $3\times3$ Higgs mixing matrices as a function of $\tan \beta$. It is important to emphasize that we keep Higgs masses and mixings as free, constrained only by the experimental results, and we do not determine them by minimizing the Higgs potential imposing the correct breaking of the electroweak symmetry. This implies that some of the considered points may not be possible to achieve in a complete model, although most of them can be reproduced with appropriate parameters in the SUSY sector. 
The main supersymmetric parameters affecting the Higgs sector, and also the indirect processes $B\to X_s \gamma$ and $B_s\to \mu^+ \mu^-$, are basically third generation masses and couplings, because of their large Yukawa couplings, and gaugino masses. In our analysis, these parameters take general values consistent with the experimental constraints on direct and indirect searches.  

In the literature there have been several works constraining the
parameter space of different MSSM variants with LHC data
\cite{Heinemeyer:2011aa,Arbey:2012dq,Bechtle:2012jw,Ke:2012zq,Ke:2012yc,Moretti:2013lya,Scopel:2013bba}
with special emphasis in light Higgs masses and the non-decoupling MSSM limit
\cite{Barenboim:2007sk,Mahmoudi:2010xp,Hagiwara:2012mga,Arganda:2012qp,Christensen:2012si,Han:2013mga,Barbieri:2013nka,Arganda:2013ve}. As described in the previous paragraph, our
analysis in a generic MSSM model includes all these MSSM variants and
updates them with the latest data on searches in the $\tau\tau$
channel from ATLAS and CMS at LHC.  Furthermore, our analytic approach with 
a few key
phenomenological observables, the two photon signal strength, the
$\tau\tau$ production cross sections at LHC and the indirect
constraints on BR$(B\to X_s \gamma)$, can neatly exclude wide regions
of the parameter space without the risk of missing a small region
where unexpected cancellations or combinations can take place and simultaneously allows us to identify clearly the observables responsible of this exclusion.

This work is organized as follows. We begin by describing the basic
ingredients of the model in Section~\ref{sec:model} and recount the latest 
results on extra-Higgs searches in Section~\ref{sec:extra-Higgs}. In
section~\ref{sec:analysis} we analyze the present constraints on the
model and the future prospects for the searches of additional Higgs
states. We discuss the possibility of a second peak in the diphoton spectrum in Section~\ref{sec:second}. Finally, results and conclusions are summarized in
Section~\ref{sec:conclusions}.
  
\section{Higgs sector in a complex MSSM}
\label{sec:model}

As explained above, we aim to establish the identity of the observed
scalar resonance found at $m_H\simeq 126$~GeV in LHC experiments and,
in particular, to check whether this is one of the Higgs states in an
MSSM setup. The MSSM is the most simple supersymmetric extension of
the SM. In this work, we carry our analysis in a generic MSSM in
the presence of CP-violating phases. Even though the Higgs sector is
CP-conserving at tree-level  \cite{Djouadi:2005gj}, the presence of 
CP-violating phases in
the theory induces at loop level CP violation in the Higgs
potential \cite{Pilaftsis:1998dd,Pilaftsis:1998pe,Pilaftsis:1999qt,Demir:1999hj,Carena:2000yi,Choi:2000wz,Carena:2001fw,Choi:2001pg,Choi:2002zp}. These loop corrections produce a mixing between scalar and
pseudoscalar states, turning this way the physical mass eigenstates
into admixtures of CP even and CP odd states, with no definite CP
parity.  Thus, the introduction of CP phases set us a far cry from the
CP conserving MSSM where the neutral scalars, $h^0$ and $H^0$, and the
pseudoscalar, $A^0$ do not mix.

Including CP violating phases into the MSSM requires to write the two
scalar doublets as  \cite{Pilaftsis:1999qt,Carena:2000yi,Carena:2001fw,Funakubo:2002yb},
\begin{equation}
\Phi_{1}=\left(\begin{array}{c}
\frac{1}{\sqrt{2}}\left(\upsilon_{1}+\phi_{1}+ia_{1}\right)\\\phi_{1}^{-}
\end{array}\right);\;\;\Phi_{2}=e^{i\xi}\left(\begin{array}{c}
\phi_{2}^{+}\\
\frac{1}{\sqrt{2}}\left(\upsilon_{2}+\phi_{2}+ia_{2}\right)
\end{array}\right)\,,\label{eq:3.1-2}
\end{equation}
where $\upsilon_{1}$ and
$\upsilon_{2}$ are the Higgs vacuum expectation values in the
electroweak vacuum  and $\tan\beta=\upsilon_{2}/\upsilon_{1}$. Now, 
the mass matrix for the physical neutral scalars, in the basis, 
$\left(\phi_{1},\phi_{2},a\right)$ with $a=a_{1}\sin\beta+a_{2}\cos\beta$, 
becomes 
\begin{equation}
M_{H}^{2}=\left(\begin{array}{cc}
M_{S}^{2} & M_{SP}^{2}\\
M_{PS}^{2} & M_{P}^{2}
\end{array}\right)\,,
\end{equation}
 where $M_{S}^{2}$ is $2\times2$ and $M_{SP}^{2} = (M_{PS}^{2})^T$ is a $1\times2$ block. This mass matrix is diagonalized by a $3\times 3$ matrix, ${\cal U}$,
\begin{equation}
{\cal U}\cdot M_{H}^{2}\cdot{\cal U}^{T}=\mbox{Diag}\left(m_{H_{1}}^{2},m_{H_{2}}^{2},m_{H_{3}}^{2}\right)\,.
\end{equation}
The scalar-pseudoscalar mixing, which is absent in the CP conserving 
case, arises at the one-loop level in the CP violating MSSM and is of order \cite{Pilaftsis:1999qt},
\begin{equation}
\label{eq:scalar-pseudo}
M_{SP}^2 = O\left(\frac{m_t^4 |\mu| |A_t|}{32 \pi^2~\upsilon^2 M_{SUSY}^2}\right) \sin\phi_{CP} \times \left[6, \frac{|A_t|^2}{M_{SUSY}^2}, \frac{|\mu|^2}{\tan\beta M_{SUSY}^2}\right]\,,
\end{equation}
where $\phi_{CP} = \arg(\mu A_{t,b}e^{i\xi})$ and  $M_{SUSY}^2 =(m_{\tilde t_1}^2 + m_{\tilde t_2}^2)/2$. 
From this expression, we see that large effects in the Higgs sector due to the presence of this CP violating phase are obtained for $\mbox{Im}\left[\mu A_{t,b}e^{i\xi}\right] \gtrsim M_{SUSY}^2$ and $M^2_P$ not much larger than $\upsilon^2$. This situation is still possible phenomenologically outside the decoupling limit and thus in the following we analyze this complex MSSM which, obviously, includes the usual real scalar potential as a particular case.  

In our analysis, we consider a generic MSSM defined at the electroweak
scale with the lightest Higgs mass $m_{H_1}\simeq 126$~GeV. We take the other 
two neutral Higgses and the charged Higgs masses as free with generic mixing 
matrices ${\cal U}$, which we constrain with the present experimental
results. The rest of MSSM parameters are also free and independent at 
$M_{W}$ and only constrained by experimental results without further theoretical restrictions. Nevertheless, some of the parameters of the Higgs sector of
the MSSM, and in particular the $\mu$ term in the superpotential, are
very important in other sectors of the theory like sfermion masses and
left--right mixings and play a very important role in several of
the analyzed flavour changing decays. In this work, we are not fixing the
value of the $\mu$ term through the requirement of correct electroweak
symmetry breaking, but we take it to vary in a wide range from
$M_2 \leq \mu \leq 3 m_{H^\pm}$, taking into account that,
at tree-level, the scale of the charged and heavy Higgses is fixed by
the $\mu$ parameter.

The remaining SUSY masses and mixings are fixed by the SUSY soft
breaking terms and are only subject to the
experimental constraints from direct LHC searches and contributions to
FCNC, as summarized in \cite{Lee:2003nta,Barenboim:2013bla} \footnote{In particular, we allow the trilinear couplings $A_i$ to take values in the range $0 \leq A_i \leq 3 m_{\tilde i}$, to avoid charge and color-breaking minima. }. In our analysis of the
Higgs sector and FCNC constraints, the most important SUSY parameters
are gaugino masses and third generation sfermion masses and mixings. 
The complete
expressions for Higgs production and decays taking into account the
couplings of the new Higgs states to fermions, scalars and gauge
bosons, can be found in \cite{Barenboim:2013bla}. 

\section{Extra-Higgs searches in the $pp\to\tau\tau$ process}
\label{sec:extra-Higgs}

The experimental constraints we impose on the Higgs sector of our generic MSSM were already described in our previous paper, \cite{Barenboim:2013bla}. The $H\to \gamma \gamma$ bounds and the indirect constraints, BR$(B\to X_s \gamma)$ and  BR$(B_s\to \mu \mu)$, are still the same and we refer to  \cite{Barenboim:2013bla} for details.

Still, the results in the channel  $H_i\rightarrow\tau\tau$, which play a very important role in the searches for additional heavy Higgs states, has been recently updated by the CMS collaboration \cite{CMS:2013hja}. Both ATLAS and CMS 
experiments had previoulsy carried dedicated analysis in this channel.
Both experiments have searched for the SM Higgs boson
decaying into a pair of $\tau$-leptons and this provides a limit
on $\sigma(pp\to H)\times\mbox{BR}(H\to\tau\tau)$ at 95\% C.L. that can be applied
to all three neutral Higgs states in the MSSM. ATLAS has analyzed the collected data samples
of $4.6\,\mbox{fb}^{-1}$ at $\sqrt{s}=$7 TeV and $13.0\,\mbox{fb}^{-1}$ at
$\sqrt{s}=$8 TeV \cite{Aad:2012mea} while CMS used $4.9\,\mbox{fb}^{-1}$ at $\sqrt{s}=$7 TeV and $19.4\,\mbox{fb}^{-1}$at $\sqrt{s}=$8 TeV for Higgs masses
up to 150 GeV \cite{CMS-PAS-HIG-13-004}. For this range of masses, CMS sets
the strongest bound: for $m_{H}=110$ GeV
we obtain a bound at 95\% CL of $\mu_{\tau\tau}=\sigma\left(H\rightarrow\tau\tau\right)/\sigma_{SM}\leq1.8$, and this limit remains nearly constant, $\mu_{\tau\tau}\leq2.0$, up to $m_{H}=140$~GeV. For a neutral Higgs of mass $m_{H}=150$~GeV we would have a
bound of $\mu_{\tau\tau}\leq2.3$. In our generic MSSM, this limit would apply to the lightest Higgs with $m_{H_1}\simeq 126$~GeV and to the two heavier neutral Higgs states when their masses are below 150 GeV.

\begin{figure}[t]
\noindent \begin{centering}
\includegraphics[scale=0.45]{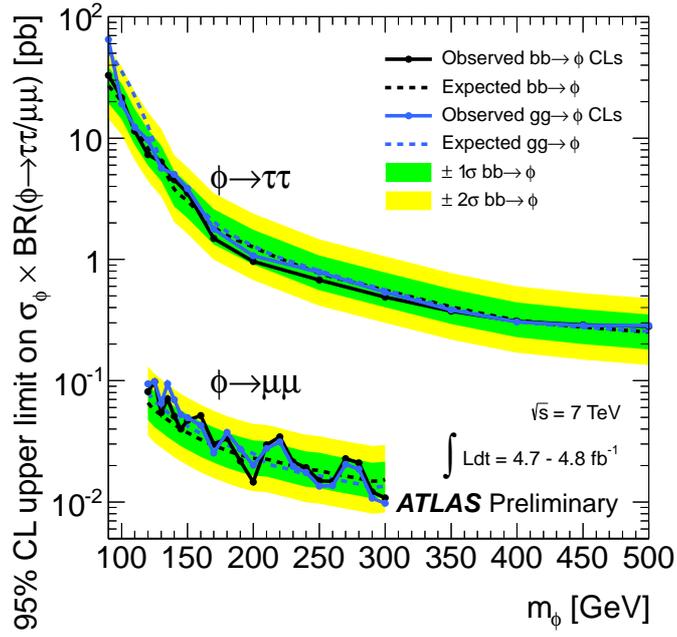}\caption{Upper limit on the $\tau\tau$ production cross section through heavy
Higgs states from ATLAS with $4.8~ \mbox{fb}^{-1}$ at $\sqrt{s}=7$~TeV \label{fig:ATLAS-MSSM-H}.}
\par\end{centering}
\end{figure}

\begin{figure}[t]
\noindent \begin{centering}
\includegraphics[scale=0.36]{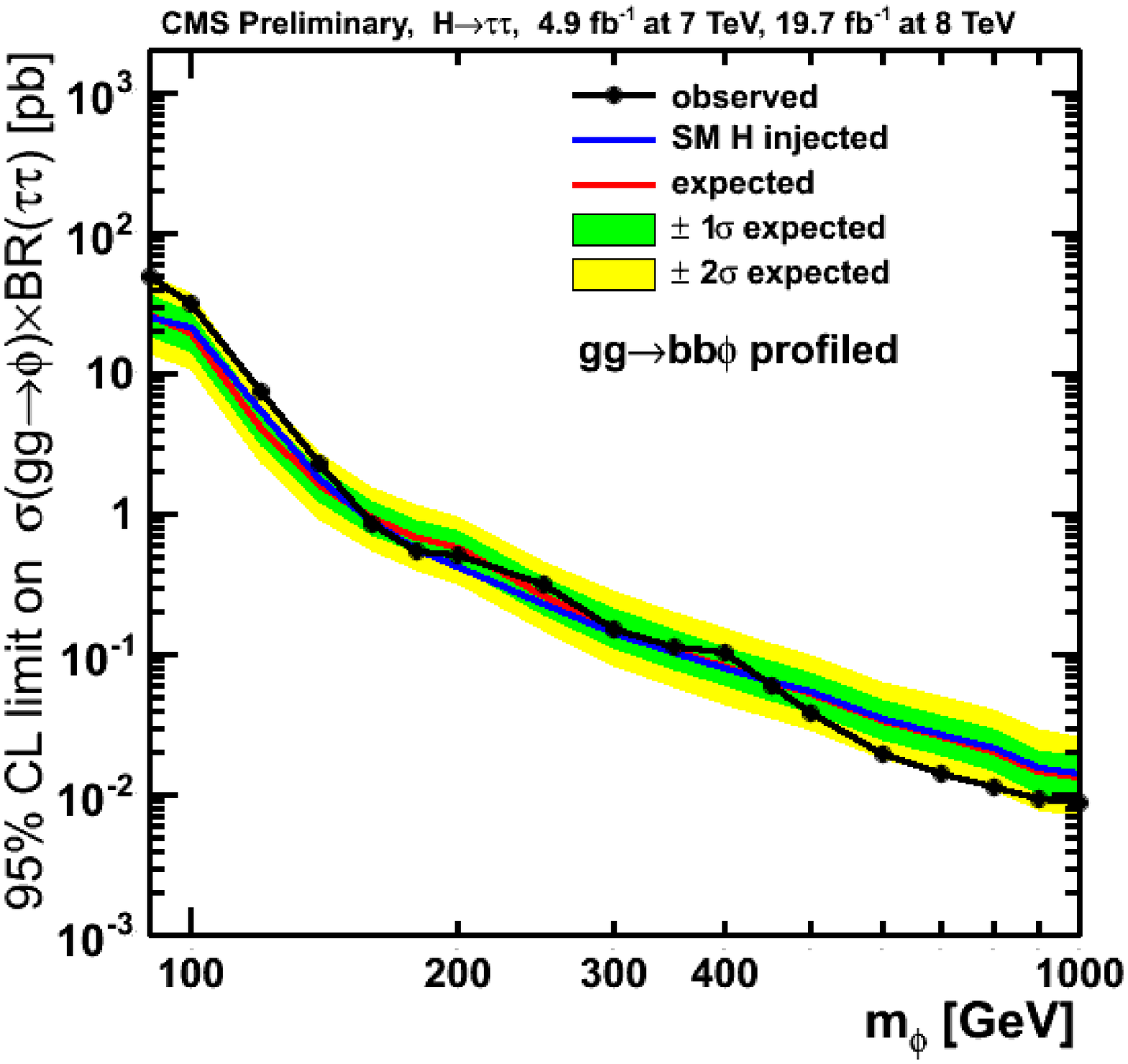}~\includegraphics[scale=0.36]{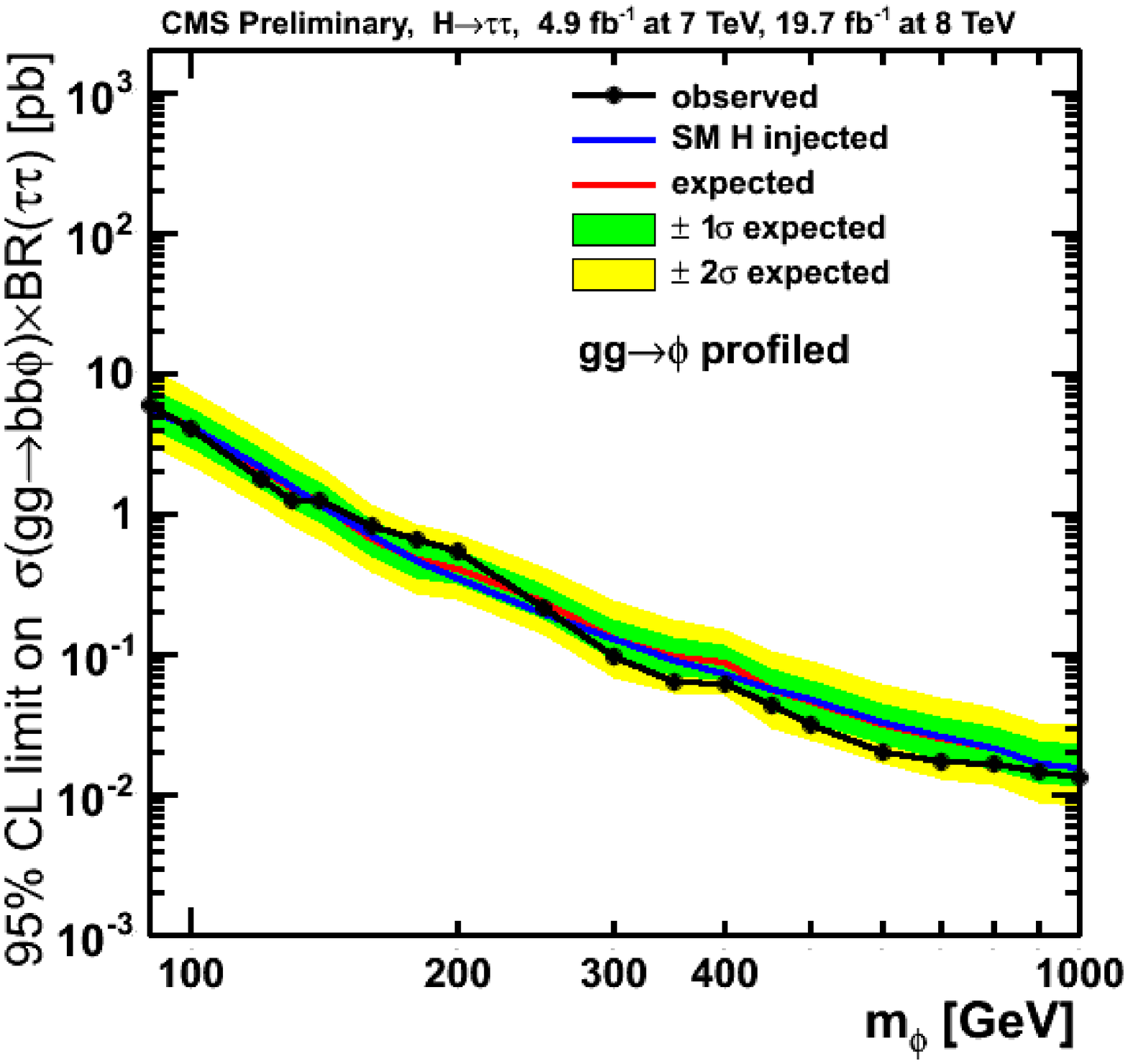}\caption{Latest CMS results on the $\tau\tau$ production cross section through heavy
Higgs states with $24.6~ \mbox{fb}^{-1}$ at $\sqrt{s}$=7--8~TeV. On the left the bound obtained from gluon-fusion produced Higgs while on the right the bound from the  $b \bar b$ production mode with two additional $b$-jets is shown. \label{fig:CMS-MSSM-H}}
\par\end{centering}
\end{figure}

For heavier $H_i$ masses, there exist a previous ATLAS analysis at LHC searching
MSSM Higgs bosons with masses up to 500 GeV  with $4.9\,\mbox{fb}^{-1}$ at
$\sqrt{s}=$7 TeV \cite{Aad:2012cfr}.  In this case, the bound as an upper limit on the $\tau\tau$, or $\mu\mu$ production cross section also at 95 \% C.L. that is shown in Figure \ref{fig:ATLAS-MSSM-H}. We can expect this bound to improve nearly an order of magnitude in an updated analysis with the new data \cite{privateFiorini}. Nevertheless, recently the CMS collaboration has presented an analysis of the full data set with an integrated luminosity of 24.6 fb$^{-1}$, with 4.9 fb$^{-1}$ at 7 TeV and 19.7 fb$^{-1}$ at 8 TeV searching for additional neutral Higgs states in the $\tau \tau$ channel up to masses of 1 TeV \cite{CMS:2013hja}. The analysis discriminates between Higgses produced through gluon fusion and $b \bar b$ fusion with two extra b-jets.
These latest CMS results are presented in Figure~\ref{fig:CMS-MSSM-H}. As we will see later, these new experimental results set very stringent constraints for the neutral Higgs spectrum. In the following, we apply all these bounds at 95 \% C.L. on the theoretical cross sections obtained in our generic MSSM.

\section{Model analysis}
\label{sec:analysis}
The purpose of this section is to present the outcome of our complex MSSM model 
predictions in different Higgs decay channels and compare them to the current
experimental results at LHC. The main Higgs search channels that we use to constrain the parameters in our model are
$pp\rightarrow H_{1}\rightarrow\gamma\gamma$ and
$pp\rightarrow H_{i}\rightarrow\tau^+\tau^-$. However, we will see that 
indirect new 
physics searches also play a very important role in constraining the model 
due the charged and neutral Higgs contributions to $b\rightarrow s\gamma$
and $B_{s}\rightarrow\mu^{+}\mu^{-}$.

In the following, after setting the lightest neutral Higgs mass at
$m_{H_{1}}=126\,\mbox{ GeV}$, we impose the constraints derived from the LHC results on $pp\rightarrow H_{1}\rightarrow\gamma\gamma$ and indirect bounds from low energy experiments. Then, we divide our 
analysis in two different regions to study the $\tau\tau$ production cross section: i) a light MSSM Higgs sector, defined by $m_{H^{+}}<m_{t}$,
that can be considered the non-decoupling regime, and ii) heavy Higgs
masses, when $m_{H^{+}} > m_{t}$, as would correspond to the decoupling
limit in the sense discussed below Eq.~(\ref{eq:scalar-pseudo}) of 
$M_P^2 > \upsilon^2$.
\subsection{Two photon cross section}

The decay $H\to \gamma\gamma$ has been the main channel in the
discovery of a scalar resonance at  $m_H \simeq 126$~GeV at LHC experiments. ATLAS finds an excess of  2.8 $\sigma$ local significance at a mass of $m_H =126.5$~GeV while CMS finds a contribution at a mass of $m_H =126.5$~GeV with a p-value of 4.1 $\sigma$. The measured signal
strength in this channel, defined as the ratio of the measured $\gamma
\gamma$ production cross section to the SM expected value, combining
both ATLAS and CMS results at two sigma is
$0.75\leq\mu_{\gamma\gamma}^{\mathrm{LHC}}\leq1.55$ and we impose the accepted 
points in the parameter space to be within this range.  
The $\gamma \gamma$ production through the Higgs in the narrow width 
approximation depends on the Higgs
production cross section and the $H \to \gamma \gamma$ branching
ratio, which in turn depends both on the decay width into two photons
and on the total decay width. Thus, we will have to analyze these three 
elements to constrain our model. 

First of all, we focus on the Higgs
decay amplitude into photons,
$\Gamma\left(H_{1}\rightarrow\gamma\gamma\right)$, which has both
scalar and pseudoscalar amplitudes that receive contributions from
gauge bosons, fermions, sfermions and charged Higgs in our MSSM model, {\it i.e.},
\begin{equation}
\Gamma\left(H_{a}\rightarrow\gamma\gamma\right)=\frac{M_{H_{a}}^{3}\alpha^{2}}{256\pi^{3}\upsilon^{2}}\left[\left|S_{a}^{\gamma}\left(M_{H_{a}}\right)\right|^{2}+\left|P_{a}^{\gamma}\left(M_{H_{a}}\right)\right|^{2}\right]\,.\label{eq:gammawidth}
\end{equation}
The full expressions for the different contributions to the scalar $S_{a}^{\gamma}$, and pseudoscalar, $P_{a}^{\gamma}$, amplitudes are presented in \cite{Barenboim:2013bla}. 
The dominant contributions to the scalar amplitude are given by the $W$-boson and 
top quark, with the bottom-quark contributing only for very large $\tan \beta$,
\begin{eqnarray}
\label{diphotonSM}
S_{H_{1}^{0},W}^{\gamma} &\simeq&-8.3\,\left(\mathcal{U}_{12}+\frac{\mathcal{U}_{11}}{\tan\beta}\right) \\
S_{H_{1}^{0},b+t}^{\gamma}&\simeq&1.8~\mathcal{U}_{12} +\left(-0.025+i\,0.034\right)\left[\mbox{Re}\left\{ \frac{\tan\beta}{1+\kappa_{d}\tan\beta}\right\} \,\mathcal{U}_{11}+ 
\mbox{Im}\left\{ \frac{\kappa_{d}\tan^{2}\beta}{1+\kappa_{d}\tan\beta}\right\} {\cal U}_{13}\right]\,,\nonumber
\end{eqnarray}
where $\kappa_d =(\Delta h_{d}/h_{d})/(1+\delta h_{d}/h_{d})$ encodes the loop corrections to the down Yukawas, with $(h_d +\delta h_{d})$ the one-loop corrected Yukawa coupling of down quarks to $H_1$ and $ \Delta h_{d}$ the non-holomorphic coupling of down quarks to $H_2^*$ \cite{Hall:1993gn,Carena:1994bv,Blazek:1995nv,Carena:1999py,Hamzaoui:1998nu,Babu:1999hn,Isidori:2001fv,Dedes:2002er,Buras:2002vd,Lee:2003nta,Carena:2002bb}.

Next, we have to consider the MSSM contributions to the amplitude, that include the charged Higgs and third generation sfermions. In the analysis of Ref.~\cite{Barenboim:2013bla} we showed that the charged Higgs contribution can always be neglected in comparison to the contributions in Eq.~(\ref{diphotonSM}). In the case of the stop, if we impose the LHC bound on the stop mass for large mass
difference to the LSP $m_{\tilde{t}} \gtrsim 650\,\mathrm{GeV}$, its contribution is also much smaller than the dominant SM contributions. However, this contribution can be somewhat larger for lighter stops with a small mass difference with the LSP, although there exists an absolute lower
bound of $m_{\tilde{t}} \gtrsim 250\,\mathrm{GeV}$ from single jet
searches \cite{Delgado:2012eu}. For low stop masses, the stop contribution can be important and we keep it in our approximate expression for the scalar amplitude.

\begin{figure}
\begin{centering}
\includegraphics[scale=0.5]{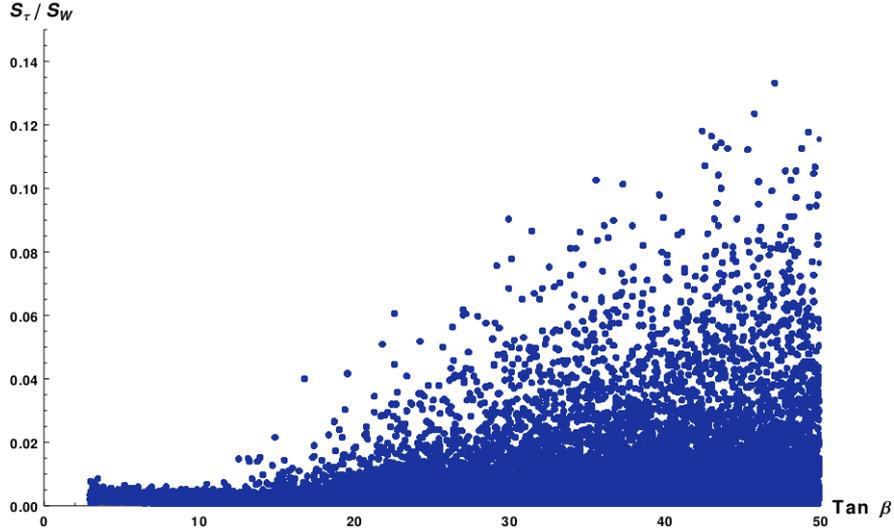}
\par\end{centering}
\caption{{\footnotesize \label{figure5}$\tilde{\tau}$ scalar contribution
to the two photons decay width compared to the and W-boson
contribution as a function of $\tan\beta$}}
\end{figure}

Finally, we have to consider sbottom and stau contributions. These
contributions are negligible at medium-low $\tan \beta$, say $\tan \beta \lesssim 8$, compared to
those coming from the SM particles due to the smallness of the Yukawa
couplings in this regime. However, they can be sizeable for very large
$\tan \beta$ or very light sparticles.  In fact, in
Refs. \cite{Carena:2011aa,Carena:2012gp,Carena:2013iba} the stau
contribution was proposed as a way to increase the diphoton decay rate
without affecting the Higgs production cross section\footnote{Notice that a large sbotton contribution would
  enhance both the Higgs production and the diphoton decay width, and
  thus modify also the successful $ZZ$ and $WW$ predictions.} and therefore not
modifying the successful predictions in other channels. However, this would 
require large $\tan \beta$ values and, as we show below, this is
incompatible with the bounds from $H_2, H_3 \to \tau \tau$ for
$m_{H_{2,3}} \leq 1$~ TeV.  Nevertheless, for such heavy Higgs masses, a light stau  could contribute considerably to the scalar amplitude for large
$\tan\beta$. The stau contribution to
$S_{H_{1}^{0}}^{\gamma}$ can be approximated by
\begin{eqnarray}
S_{H_{1}^{0},\,\tilde{\tau}}^{\gamma}&\simeq&0.36\tan^{2}\beta\frac{m_{\tau}^{2}}{m_{\tilde{\tau}_{1}}^{2}}\left[\frac{\mathrm{Re}\left\{
    A_{\tau}^{*}\mu\right\}
  }{m_{\tilde{\tau}_{2}}^{2}}\mathcal{U}_{11}\text{\textSFx}\frac{\mu^{2}}{m_{\tilde{\tau}_{2}}^{2}}\mathcal{U}_{12}+\frac{\mathrm{Im}\left\{
    A_{\tau}^{*}\mu\right\}
  }{m_{\tilde{\tau}_{2}}^{2}\tan\beta}\mathcal{U}_{13}\right]\\&\simeq&3
\times 10^{-5} \tan^{2}\beta \left( \frac{200 {\rm GeV}}{m_{\tilde
    \tau_1}} \right)^2\left[\frac{\mathrm{Re}\left\{
    A_{\tau}^{*}\mu\right\}
  }{m_{\tilde{\tau}_{2}}^{2}}\mathcal{U}_{11}\text{\textSFx}\frac{\mu^{2}}{m_{\tilde{\tau}_{2}}^{2}}\mathcal{U}_{12}+\frac{\mathrm{Im}\left\{
    A_{\tau}^{*}\mu\right\}
  }{m_{\tilde{\tau}_{2}}^{2}\tan\beta}\mathcal{U}_{13}\right]\, ,
\nonumber
\end{eqnarray}
where we used that the loop function is approximately 0.35 (and tends to 1/3) for $m_{\tilde{\tau}} \gtrsim 200\,\mathrm{GeV}$. From here it is clear that an $O(1)$ stau contribution, which would be required to enhance the diphoton rate, is only possible for $\tan \beta \geq 80$ and  $m_{\tilde{\tau}} \leq 100$~GeV if $ A_{\tau}^{*}/m_{\tilde \tau_2},\mu/m_{\tilde \tau_2} \simeq O(1)$. Even in an extreme case, $ A_{\tau}^{*}/m_{\tilde \tau_2},\mu/m_{\tilde \tau_2} \lesssim 3$, would require $\tan \beta \geq 50$ and $m_{\tilde{\tau}} \leq 150$~GeV to get an $O(1)$ contribution.  This can be seen in Figure \ref{figure5}, where we compare the stau and the W-boson contributions up to $\tan \beta$ values of 50.
In any case, these large $\tan \beta$ values can increase the diphoton width at most a 10--30\% and, as we show below, such large $\tan \beta \gtrsim 50$ values are strongly constrained by $H_i \to \tau \tau$ channel. Moreover, this large $\tan \beta$ values would not be enough to increase the $H_1 \to \gamma \gamma$ branching ratio, as they would simultaneously increase the $H_1$ total width. Therefore, in this work, we do not consider such large $\tan \beta$ values and we neglect stau and sbottom contributions.  Also, as we show in~\cite{Barenboim:2013bla}, chargino contributions are always negligible.  

Thus, finally, keeping only the dominant W-boson, quark and stop contributions, we can approximate the scalar amplitude by,
\begin{eqnarray}
\label{eq:gammaampl}
S_{H_{1}^{0}}^{\gamma}&\simeq&\mathcal{U}_{11}\,\left(-\frac{8.3}{\tan\beta} + \left(-0.025+i\,0.034\right)\,\mbox{Re}\left\{ \frac{\tan\beta}{1+\kappa_{d}\tan\beta}\right\}  \right.\\&
 & \left.\quad\qquad
- 0.45\,\left(\frac{m_{\tilde{t}_{2}}^{2}}{m_{\tilde{t}_{1}}^{2}}-1\right)\mbox{Re}\left\{ \frac{\mu m_{t}\mathcal{R}_{11}^{*}\mathcal{R}_{21}}{m_{\tilde{t}_{2}}^{2}}\right\}\right)\, +\nonumber \\&&\mathcal{U}_{12}\, \left(-6.5+0.45\,\left(\frac{m_{\tilde{t}_{2}}^{2}}{m_{\tilde{t}_{1}}^{2}}-1\right)\mbox{Re}\left\{ \frac{A_{t}^{*}m_{t}\mathcal{R}_{11}^{*}\mathcal{R}_{21}}{m_{\tilde{t}_{2}}^{2}}\right\} \right.\nonumber \\&&\left.\quad\qquad+0.45\,\left(\frac{m_{t}^{2}\left|\mathcal{R}_{11}\right|^{2}}{m_{\tilde{t}_{1}}^{2}}+\frac{m_{t}^{2}\left|\mathcal{R}_{12}\right|^{2}}{m_{\tilde{t}_{2}}^{2}}\right)\right)+\nonumber\\
& & {\cal U}_{13}\, \left(\left(-0.025+i\,0.034\right)\,\mbox{Im}\left\{ \frac{\kappa_{d}\tan^{2}\beta}{1+\kappa_{d}\tan\beta}\right\} \right.\nonumber \\&&\left.\quad\qquad +0.45\,\left(\frac{m_{\tilde{t}_{2}}^{2}}{m_{\tilde{t}_{1}}^{2}}-1\right)\mbox{Im}\left\{ \frac{\mu m_{t}\mathcal{R}_{11}^{*}\mathcal{R}_{21}}{m_{\tilde{t}_{2}}^{2}}\right\} \right)\,.\nonumber 
\end{eqnarray}
This amplitude has to be compared with the SM value $S_{H_{SM}}^{\gamma}\simeq -6.55$. The pseudoscalar amplitude, absent in the SM, is typically much smaller, as it receives contributions only from fermions, i.e. mainly top and bottom quarks, and these contributions are of the same order as fermionic contributions to the scalar amplitude.  
 
Then, the total Higgs decay width receives contributions mainly from
$H_{1}\rightarrow WW^{*}$ and the down-type fermion, $H_{1}\rightarrow b\bar{b}$ and $H_{1}\rightarrow \tau\tau$ which, compared to the 
SM predictions, are enhanced by $\tan^{2}\beta$.
Then $H_{1}\rightarrow gg$ decay can be of the same order of $H_{1}\rightarrow \tau\tau$ for low $\tan \beta$, but can be safely neglected as it is always subdominant with respect to $b\bar b$ and $W W^*$ and does not influence significantly the total width:
\begin{equation}
\Gamma_{H_{1}}\simeq\frac{g^{2}m_{H_{1}}}{32\pi M_{W}^{2}}\left[\tan^{2}\beta\left(\mathcal{U}_{11}^{2}+\mathcal{U}_{13}^{2}\right)\left(3m_{b}^{2}+m_{\tau}^{2}\right)+I_{PS}\left(\mathcal{U}_{12}+\frac{\mathcal{U}_{11}}{\tan\beta}\right)^{2}m_{H_{1}}^{2}\right]\label{eq:total width}
\end{equation}
with  $I_{PS}\simeq6.7\times10^{-4}$ being the phase space integral.

Using Eqs.~(\ref{eq:gammawidth}) and (\ref{eq:total width}) we can estimate BR($H_1 \to \gamma \gamma$) as,
\begin{eqnarray}
{\rm BR}(H_1 \to \gamma\gamma)&\simeq&\frac{\alpha^2}{32 \pi^2 \left(3x_{b}+x_{\tau}\right) }~\frac{|S^\gamma|^2 + |P^\gamma|^2}{\left(\mathcal{U}_{11}^{2}+\mathcal{U}_{13}^{2}\right)\tan^{2}\beta+\left(\mathcal{U}_{12}+\frac{\mathcal{U}_{11}}{\tan\beta}\right)^{2}~\frac{I_{PS}}{\left(3x_{b}+x_{\tau}\right)}}\nonumber \\&\simeq& 4.65 \times 10^{-3} ~\frac{|S^\gamma/6.5|^2 + |P^\gamma/6.5|^2}{\left(\mathcal{U}_{11}^{2}+\mathcal{U}_{13}^{2}\right)\tan^{2}\beta+ 0.38~\left(\mathcal{U}_{12}+\frac{\mathcal{U}_{11}}{\tan\beta}\right)^{2}} \, .
\end{eqnarray}
From here we can see that it is very difficult to obtain a diphoton branching ratio larger than the SM value, $\sim 3 \times 10^{-3}$. In fact, the branching ratio is inversely proportional to $\tan^2 \beta$ for $\mathcal{U}_{11}\sim O(1)$, and from the diphoton decay width, Eq.~(\ref{eq:gammaampl}), we see that there is no way to compensate this enhancement in the total width through a $\tan \beta$-enhanced contribution or through the stop contribution to $S^\gamma$ in the numerator consistently with present bounds on sfermion masses \cite{Barenboim:2013bla}.

Finally, the last ingredient we need is the Higgs production cross section. This cross section is dominated by gluon
fusion and $b\bar{b}$--fusion (a complete derivation can be found
in \cite{Barenboim:2013bla}). As before, the Higgs mixings
and $\tan\beta$ are the main parameters determining the
final result. The partonic tree-level $b\bar{b}$--fusion cross section together with the  $b\bar b$ luminosity of the 5-flavour MSTW2008 parton distributions  \cite{Martin:2009iq} give a $pp$ cross section of the form:
\begin{eqnarray}
\sigma\left(pp\rightarrow H_{1}\right)_{b\bar{b}}\simeq0.16\frac{\tan^{2}\beta}{\left(1+\kappa_{d}\tan\beta\right)^{2}}\left(\left|\mathcal{U}_{11}\right|^{2}+\left|\mathcal{U}_{13}\right|^{2}\right)\,\mathrm{pb}\label{eq:bbfusion}\, .
\end{eqnarray}

Whereas the gluon fusion contribution, with the gluon luminosity from MSTW2008, will be:
\begin{eqnarray}
\sigma\left(pp\rightarrow H_{1}\right)_{gg}&\simeq\left[13~\mathcal{U}_{12}^{2}+\frac{0.1\tan^{2}\beta}{\left(1+\kappa_{d}\tan\beta\right)^{2}}~\mathcal{U}_{11}^{2}-\frac{1.4\tan\beta}{1+\kappa_{d}\tan\beta}~\mathcal{U}_{11}\mathcal{U}_{12}\, + \right. \nonumber \\
&\left. \left(\frac{2}{\left(1+\kappa_{d}\tan\beta\right)}+\frac{0.1 \tan^2\beta}{\left(1+\kappa_{d}\tan\beta\right)^2}+\frac{27}{\tan^2\beta}\right)\,{\cal U}_{23}^2\right]\,\mathrm{pb}\, ,
\end{eqnarray}
where we can see that the top quark contribution is the most relevant one in the gluon fusion amplitude, except for large $\tan \beta$ and $\mathcal{U}_{11},\,\mathcal{U}_{13}\sim O(1)$ where bottom fusion and the bottom contributions to gluon fusion become important and overcome the top contribution. Nevertheless, we must keep the Higgs production cross section close to the SM values, as this is required by the  experimental results in other Higgs search channels.

\begin{figure}[h]
\begin{centering}
  \includegraphics[scale=1.2]{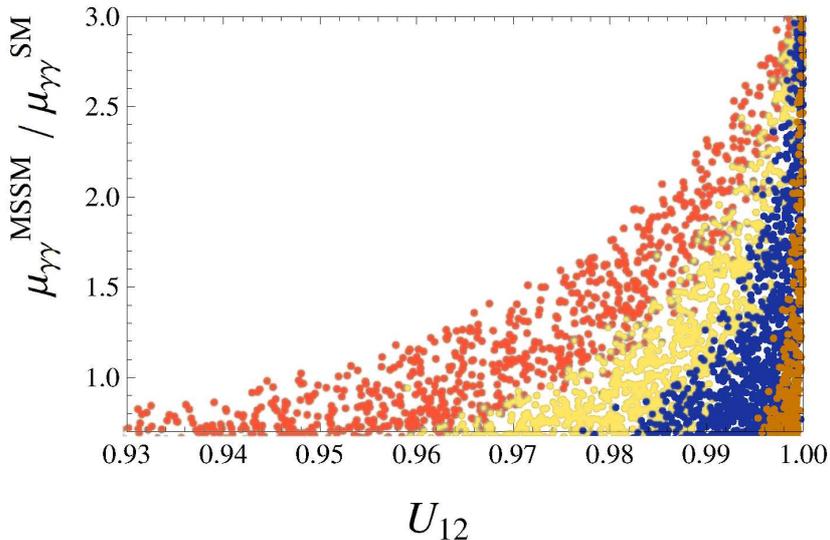}
\par\end{centering}
\caption{\label{fig:component-photons}Number of events (normalized to SM)
in $H_{1}\rightarrow\gamma\gamma$ respect to the Higgs up mixing
component in a generic MSSM as described in the text.}
\end{figure}

In summary, we have seen that the production cross section must be similar to the SM one, while the total decay width is larger than the
SM one if $\mathcal{U}_{11},\,\mathcal{U}_{13}>\tan^{-1}\beta$. Thus, we would need 
 $\mathcal{U}_{11},\,\mathcal{U}_{13}\lesssim \tan^{-1}\beta$ to reduce the total width and increase ${\rm BR}(H_{1}\rightarrow\gamma\gamma)$ to keep it of the order of the SM value. On the other hand, for small $\mathcal{U}_{11},\,\mathcal{U}_{13}$, the Higgs production is dominated by gluon
fusion and then it is possible to reproduce the observed signal strength in the different Higgs decay channels. 
Therefore, the following Higgs mixing components
appear naturally as a consequence of enlarging the value of $\mathrm{BR}\left(H_{1}\rightarrow\gamma\gamma\right)$:
\begin{equation}
\mathcal{U}_{12}\simeq1;\;\mathcal{U}_{11},\,\mathcal{U}_{13}<\frac{1}{\tan\beta}
\end{equation}
In Figure \ref{fig:component-photons} we show the allowed $U_{12}$ values as a function of the diphoton signal strength. The different colours correspond to different $\tan \beta$ values with $\tan \beta < 5$ orange,  $5<\tan \beta < 9$ yellow,  $9<\tan \beta < 30$ blue and  $30<\tan \beta$ brown. From here, it is clear that $\mathcal{U}_{12}$ is required to be close to one, as $1 - (1/\tan \beta)^2$.  Notice that this result simply generalizes the usual real MSSM result in the decoupling limit, which implies that $\mathcal{U}_{12}=\cos \alpha\simeq \sin \beta$.
Once our model satisfies the requirement of the observed signal strength in the diphoton channel, we impose next the limits on $H_a \to \tau \tau$ and BR($B \to  X_s \gamma$). As we will see, the first and second constraints are more relevant for medium--high or low $\tan \beta$ values respectively.  Then, we divide the analysis in two areas depending on the masses of the extra Higgses: i) light Higgs masses, $m_{H^\pm}\leq m_t$, where we have strong constraints from the experimental searches of the SM Higgs and ii) heavy Higgs masses, $m_{H^\pm}\geq m_t$, where at the moment, we can use the searches of MSSM neutral Higgses with 4.8 fb$^{-1}$ from ATLAS \cite{Aad:2012cfr} and 24.6 fb$^{-1}$ from CMS experiment \cite{CMS:2013hja}.  

\subsection{$H_a \to \tau \tau$ production cross section}

The $pp\to H\to \tau \tau$ production cross section is one of the main channels used to search for Higgs boson states at LHC. For Higgs masses below 150 GeV,
ATLAS and CMS put stringent bounds on this cross section with
13.0~fb$^{-1}$ and 19.4~fb$^{-1}$ respectively at $\sqrt{s} =
8$~TeV. Larger Higgs masses are constrained by the search for MSSM
Higgs bosons at ATLAS, but only with 4.8~fb$^{-1}$ at $\sqrt{s} = 7$~TeV \cite{Aad:2012cfr} .
Recently the CMS colaboration has released a complete analysis with the full collected datat set at $\sqrt{s} = 7$ and 8~TeV, which looks for extra neutral Higgses with masses up to 1~TeV \cite{CMS:2013hja}.

As we have seen, the lightest Higgs with $m_{H_1}= 126$~GeV must be
mainly up-type to reproduce the observed signal strength. Thus, the
$\tan \beta$ enhancement of the decay width of $H_1$ into tau fermions
is controlled by this small mixing. However, for the heavier neutral Higgses,
we have the opposite effect and the down-type or pseudoscalar content of
the heavier Higgses is high and, thus, the $H_{2,3}\to \tau^+\tau^-$ decay width will be, 
 at tree level and neglecting the relatively small non-holomorphic corrections to the tau Yukawa,
proportional to $\tan^{2}\beta$ in the form:
\begin{equation}
\Gamma_{i,\,\tau\tau}\simeq\frac{g^{2}m_{H_{i}}m_{\tau}^{2}}{32\pi M_{W}^{2}}
\tan^{2}\beta\,.
\end{equation}
Here, we have to remember that the relevant quantity in the $p p \to \tau\tau$ production cross section is the $H_i$ branching ratio to $\tau^+ \tau^-$, and in this case, due to similar $\tan \beta$ enhancement of the dominant decay width into the $b \bar b$ channel, it will be basically independent of $\tan \beta$.

On the other hand, for medium--large $\tan \beta$, the
production of these Higgs bosons will also be mainly due to
$b\bar{b}$--fusion and the $b\bar{b}$ contribution to the gluon-fusion loop\footnote{In our numerical analysis, we include also  the gluon-$b$ production channel, although it is always subdominant if $b$-jets are not tagged.} and can be approximated 
by:
\begin{eqnarray}
\sigma(pp\to H_{i}) &\simeq& \left[0.07~\left(\frac{ \tau_{H_i}~d{\cal L}^{bb}/d\tau_{H_i}}{1000 ~\mbox{pb}}\right) +  0.04 ~\left(\frac{ \tau_{H_i}~d{\cal L}^{gg}_{LO}/d\tau_{H_i}}{1.1 \times 10^6 ~\mbox{pb}}\right)\right]\,\frac{\tan^{2}\beta}{(1+\kappa_{d}\tan\beta)^2}~\mbox{pb} \,,\nonumber \\
\label{eq:aprbbfusHifin}
\end{eqnarray}
where we have taken $U_{2,2}^2 +U_{2,3}^2 \simeq U_{3,2}^2 +U_{3,3}^2 \simeq 1$ and used the gluon and $b\bar b$ luminosities at $m_{H_1} = 150$~GeV at $\sqrt{s}=7$~TeV. Therefore, we can see that the $\tau \tau$ production cross section of $H_2$ and $H_3$ will be
\begin{eqnarray}
\label{eq:pphitauaprox}
&\sigma(pp &\overset{H_i}{\longrightarrow} \tau \tau)\lesssim \frac{\tan^{2}\beta}{8.4+10.4\kappa_{d}\tan\beta+\kappa_{d}^2\tan^2\beta}\nonumber \\ &&~\left[0.07\left(\frac{ \tau_{H_i}~d{\cal L}^{bb}/d\tau_{H_i}}{1000 ~\mbox{pb}}\right) +  0.04 \left(\frac{ \tau_{H_i}~d{\cal L}^{gg}_{LO}/d\tau_{H_i}}{1.1 \times 10^6 ~\mbox{pb}}\right)\right]~\mbox{pb} \,,
\end{eqnarray}
where we used the partonic luminosities for $m_{H_i}= 150$~GeV.   

The latest CMS constraints discriminate between Higgs bosons produced through gluon fusion and through $b \bar b$ fusion in association with two $b$-jets. A $p_T$-cut of 30 GeV is imposed in at least one $b$-jet in order to identify the $b \bar b$ origin. The theoretical production cross section with $b$-jets is obtained using the MSTW2008 pdf in the 5-flavour scheme \cite{Martin:2009iq} with the $b g \to h_i b$ cross section and a 30 GeV $p_T$ cut on the final $b$-jet. For this, we use the differential partonic cross section \cite{Campbell:2002zm}, 
\begin{eqnarray}
\frac{d \hat \sigma_{g b \to h_i b}}{dt}= - \frac{1}{s^2} \frac{\alpha_S(\mu)}{24} \left(\frac{y_b(\mu)}{\sqrt{2}}\right)^2 \frac{m_{h_i}^4+ u^2}{st}\,,
\end{eqnarray}
where $s,t,u$ are the Mandelstan variables. The total $pp$ cross section is then obtained as,
\begin{equation}
\sigma(pp\to h_{i}b)=4 ~\hat{\sigma}_{gb\rightarrow h_{i}b}~\int_{\tau}^{1}\frac{dx}{x}\, b(x,M^{2})\, g(\tau/x,M^{2})\,,
\end{equation}
where now $\tau =(p_g+p_b)^2/s$ and the factor 4 is due to the $b$-quark coming from one of the two protons and the conjugated process $g  \bar b \to h_i \bar b$.

On the other hand, the gluon fusion cross section without tagged b-jets is obtained as before.   

\subsection{Indirect bounds from $B \to X_s \gamma$}

After applying the constraints on the Higgs mixings from the $H_1 \to \gamma \gamma$ decay and the $H_{i}\rightarrow\tau^+ \tau^-$ decay, 
the most important constraint
will come now from an indirect flavour bound, $B\rightarrow X_{s}\gamma$. However, in our calculation we include other indirect constraints on additional Higgs states, as the top quark decay $t\rightarrow b\, H^{+}$ for light charged Higgs,  the $B^+ \to \tau^+ \nu$ decay and specially the rare decay $B_s \to \mu^+ \mu^-$, which plays a significant role for large $\tan \beta$. 

The $\mathrm{BR}\left(B\rightarrow X_{s}\gamma\right)$ has, as 
shown in \cite{Barenboim:2013bla}, a sizeable contribution coming from
the light charged-Higgs for low $\tan\beta$,
\begin{equation}
{\cal C}_{7,8}^{H^{\pm}}=\frac{f_{7,8}^{(1)}(y_{t})}{3\tan^{2}\beta}\,+\,\frac{f_{7,8}^{(2)}(y_{t})\, +\, \left(\Delta h_{d}/h_{d} \left( 1 + \tan\beta\right) - \delta h_{d}/h_{d} \left( 1 - \cot\beta\right)\right)\,f_{7,8}^{(2)}(y_{t}) }{1+\delta h_{d}/h_{d}+\Delta h_{d}/h_{d}\tan\beta} \label{eq:C7H}
\end{equation}
with $y_{t}=m_{t}^{2}/M_{H^{\pm}}^{2}$ and the loop functions $f^{(i)}_{7,8}(x)$ are defined in Ref. \cite{Barenboim:2013bla}.
We can see in this equation that the charged Higgs contribution at large $\tan \beta$ is given by the second term, which is only mildly dependent on
$\tan \beta$ due to the loop corrections to the b-quark mass. At low $\tan \beta$ values, $C_{7,8}^{H^\pm}$ increases due to a larger contribution from the first term and the reduction of denominator in the second term, and it can become sizeable for low $m_{H^\pm}$ values. This large charged-Higgs contribution cannot be compensated by the stop--chargino contribution. This is due to the $\tan \beta$ proportionality of this contribution and the small  $\tan \beta$ values that make this contribution too small even if we force the stop
mass into the region below $m_{\tilde{t}_{1}}=650\,\mathrm{GeV}$, still experimentally allowed for small stop--neutralino mass differences.

\subsection{Light MSSM Higgs masses}

We define the light Higgs region as $m_{H^{+}}<m_{t}$, being the
charged Higgs heavier than the neutral scalars of our model. In this
regime, Higgs states are strongly constrained by the present
experimental results, in particular by the process $pp\rightarrow
H_{i}\rightarrow\tau^+\tau^-$ which has been analyzed with
13.0~fb$^{-1}$ and 19.4~fb$^{-1}$ at $\sqrt{s} = 8$~TeV in ATLAS and
CMS respectively for Higgs masses below 150 GeV and with 4.8~fb$^{-1}$
at $\sqrt{s} = 7$~TeV by ATLAS for $ m_H\geq 150$~GeV. However, recently the CMS collaboration has presented an analysis with 4.9 fb$^{-1}$ at $\sqrt{s} = 7$~TeV and 19.7~fb$^{-1}$ at $\sqrt{s} = 8$~TeV \cite{CMS:2013hja}.  Furthermore,
such light charged-Higgs produce a rather large contribution to
flavour changing observables as the $B \to X_s \gamma$ decay and this constraint is very relevant in the low $\tan \beta$ region.

\begin{figure}
\begin{center}
\includegraphics[scale=0.80]{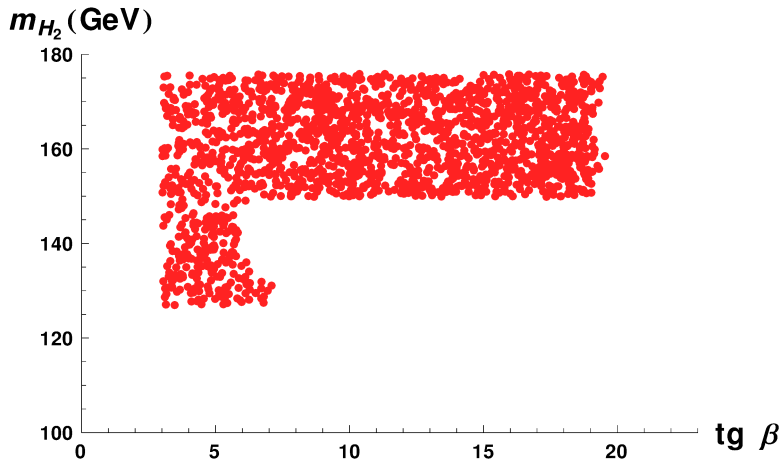}
\end{center}
\caption{\label{figure1} Allowed Higgs masses in the plane
$(\tan\beta,\: M_{H_2})$ in the light Higgs scenario. All points
satisfy the ATLAS and CMS $pp \to \tau \tau$ bounds in SM Higgs searches (for $m_{H_i}\leq 150$~GeV), plus the bounds from the ATLAS search of MSSM neutral Higgses.}
\end{figure}

Thus, the $\tau\tau$ production cross section is proportional to $\tan^2 \beta$ and we can expect the presence of additional Higgs bosons to be strongly constrained by the current searches, that are sensitive to cross sections of the order of the SM cross section for $m_H \lesssim 150$ GeV. In Figure \ref{figure1} we present the allowed Higgs masses as a function of $\tan\beta$, using only the ATLAS and CMS searches up to 150 GeV plus ATLAS MSSM Higgs searches up to 500 GeV on the channel $H_{i}\rightarrow\tau\tau$.  As we can see in this figure, these bounds eliminate completely the possibility of having additional Higgs states with masses below 145 GeV for  $\tan \beta \gtrsim 7$. This is due to the strong bounds from the SM Higgs searches in the $\tau\tau$ channel at CMS with 19 fb$^{-1}$. However, for masses $145~{\rm GeV} \leq m_{H_i} \leq 175~{\rm GeV}$, ATLAS constraints on $\sigma(pp \to H \to \tau \tau)$, shown in Figure~\ref{fig:ATLAS-MSSM-H}, are not able to exclude additional Higgs states for $\tan \beta \lesssim 24$.  

\begin{figure}
\begin{centering}
\includegraphics[scale=0.80]{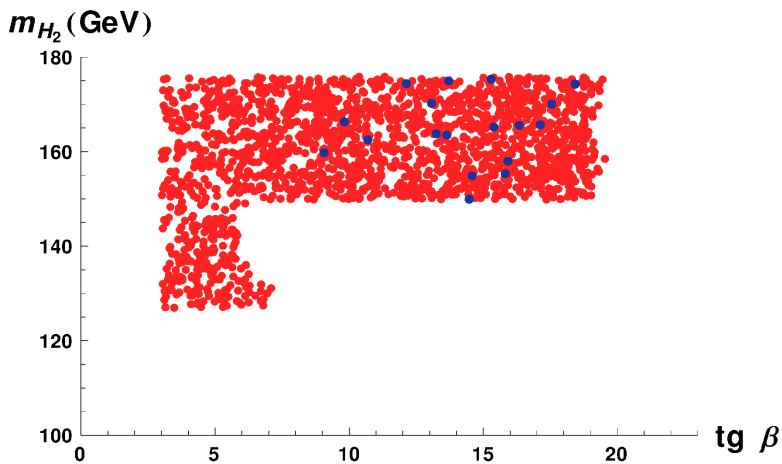}
\par\end{centering}
\caption{ \label{figure2}Allowed Higgs masses in the plane
$(\tan\beta,\: M_{H_2})$ in the light Higgs scenario. Red (dark grey) points
satisfy the ATLAS and CMS $pp \to \tau \tau$ bounds in SM Higgs searches plus ATLAS MSSM neutral Higgs searches,
whereas blue (black) points satisfy $B\rightarrow X_{s}\gamma$ in addition. 
However, if we consider improved
CMS $pp\to \tau\tau$ bounds through $b \bar b$-produced or gluon-fusion produced Higgs all the parameter space is ruled out.}
\end{figure}
Besides, if we add the constraints from the rare decay $B\rightarrow X_{s}\gamma$, most of these points are also excluded, as can be seen in Figure \ref{figure2}. All the points in this figure satisfy current $\tau \tau$ bounds, but blue points satisfy in addition $B\rightarrow X_{s}\gamma$ while red points do not satisfy this constraint. From this figure we can see that the combination of current $B\rightarrow X_{s}\gamma$ and ATLAS $\tau \tau$ bounds is able to nearly eliminate the possibility of additional Higgs states with masses below 175 GeV with the exception of a few points in the $10 \leq \tan \beta\leq 23$ range, where the charged Higgs contribution is reduced and can be compensated by a sizeable stop-chargino opposite sign contribution.

However, when the present analysis was about to be completed, the CMS collaboration relased an analysis of the full data set with 24.6 fb$^{-1}$ searching for neutral MSSM Higgs states up to 1 TeV \cite{CMS:2013hja}. In light of these results, this narrow region is completely ruled out, closing the door on the possibility of having extra Higgs states below $m_t$. 

\subsection{Heavy MSSM Higgs masses}

Next, we consider second and third neutral Higgs masses much larger than the 
lightest Higgs mass wich is fixed at the experimental value of 
$m_{H_{1}}=126\,\mathrm{GeV}$. 
In this limit, already approaching the decoupling limit in the MSSM, the heaviest mass of the scalar sector is the charged-Higgs
mass, that we take now $m_{H^{\pm}}>m_{t}$.

As we did in the previous case, we require the lightest Higgs to reproduce
the observed signal strength in the $\gamma\gamma$ channel. As we have seen, 
this implies that $H_1$ must have a dominant up-type component and therefore, 
the heavier Higgs states must be dominantly down-type or pseudoscalar. So, we can expect the $H_{2,3}\to \tau \tau$ decay width to be important. On the other 
hand, once the neutral and charged Higgs have large masses, new decay channels 
are opened, which can reduce the branching ratio of
$H_{2,3}\rightarrow\tau\bar{\tau}$. However, in the limit of large
$\tan \beta$, both the (mostly) down-type Higgs and the pseudoscalar
Higgs decay dominantly to $b \bar b $ and $\tau^+ \tau^-$ and we have that
BR($H_{2,3}\to \tau^+ \tau^-$) is typically $\sim 0.1$. 
In the low $\tan\beta$ region, and once $m_{H_i}\geq 2 m_t$, the $t \bar t$ 
channel is sizeable too and can dominate the total
Higgs width reducing in this way the $H_{2,3}\to \tau^+ \tau^-$ branching ratio. Nevertheless we will see that in this low $\tan \beta$ region, the constraints from 
$B \to X_s \gamma$ on charged Higgs masses are important and reduce significantly the allowed parameter space.

On this framework, we add now the constraints from ATLAS and CMS searches of MSSM
neutral Higgs bosons in the $\tau \tau$ channel. ATLAS searches were done 
only with 4.8~fb$^{-1}$ at $\sqrt {s} = 7$~TeV, but at the moment the collaboration has, in addition to this data, more than 20 fb$^{-1}$ at $\sqrt {s} = 8$~TeV and therefore we can expect these bounds to improve nearly an order of
magnitude in an updated analysis with the new data \cite{privateFiorini}. On the other hand, the more recent CMS analysis uses 4.9 fb$^{-1}$ at $\sqrt{s} = 7$~TeV and 19.7~fb$^{-1}$ at $\sqrt{s} = 8$~TeV and it is, at present, the key constraint on additional neutral Higgs searches.

\begin{figure}
\begin{centering}
\includegraphics[scale=0.8]{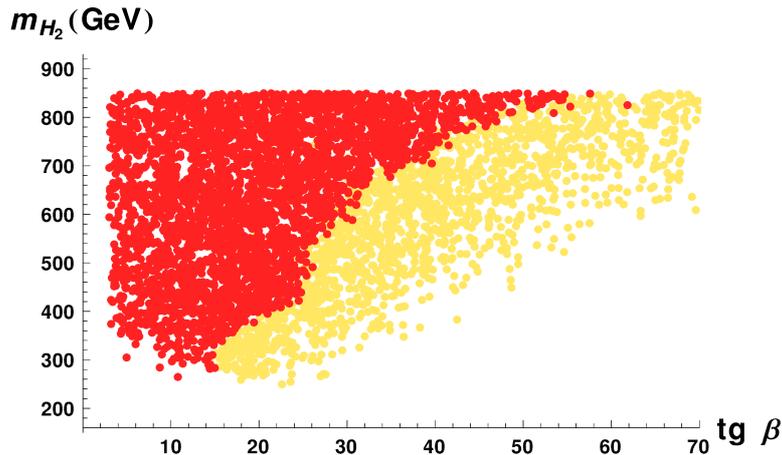}
\par\end{centering}
\caption{\label{fig:highmass}Allowed Higgs masses in the plane
$(\tan\beta,\: M_{H_2})$ taking into account the diphoton signal strength, $\tau\tau$ bounds and BR$(B \to X_s \gamma)$. 
Yellow (light grey) points are those that satisfy
the present ATLAS bounds at 95\% C.L., whereas red (dark grey) points that fulfill the 
recent CMS constraints at 95\% C.L.}
\end{figure}

As we have seen in the previous section, $\tau\tau$ constraints are very effective in the large $\tan \beta$ region. As an example, the $\tau \tau$ production cross section at $\sqrt{s}= 7$~TeV for high Higgs masses is given by,
\begin{eqnarray}
&\sigma(pp &\overset{H_i}{\longrightarrow} \tau \tau)\lesssim \frac{\tan^{2}\beta}{8.4+2\kappa_{d}\tan\beta+\kappa_{d}^2\tan^2\beta}\nonumber \\ &&~\left[0.011\left(\frac{ \tau_{H_i}~d{\cal L}^{bb}/d\tau_{H_i}}{155 ~\mbox{pb}}\right) +  0.004 \left(\frac{ \tau_{H_i}~d{\cal L}^{gg}_{LO}/d\tau_{H_i}}{1.2 \times 10^5 ~\mbox{pb}}\right)\right]~\mbox{pb} \,,
\label{eq:aprbbfusHihigh}
\end{eqnarray}
where we used the luminosities corresponding to a Higgs mass of~250
GeV. Comparing this equation with ATLAS constraints in Fig.~\ref{fig:ATLAS-MSSM-H} at $m_H
= 250$~GeV, we see that this cross section would be lower than $\sim
1.5$~pb at 95 \% CL. Thus we would obtain, from this approximate
formula and using a typical value for $\kappa_b \simeq 0.05$, a bound on $\tan \beta \lesssim 30$ at a mass $m_H = 250$~GeV. Then, we impose also the recent CMS bounds on the $pp\to H \to \tau \tau$ and  $pp\to H + b \bar b \to \tau \tau + b \bar b$ cross section. The result from these bounds is shown in Figure
\ref{fig:highmass} in the plane $(\tan\beta,\:M_{H_{2}})$. The yellow
points in this figure are allowed by ATLAS $\tau \tau$ 
constraints while red points satisfy also the stronger CMS bounds. All the points in this figure satisfy BR$(B\rightarrow X_s\gamma)$ bounds and other indirect constraints, as $B\rightarrow\tau\nu$ and
$B_{s}\rightarrow\mu^{+}\mu^{\text{\textSFx}}$. 

Indeed, we see that the combination of direct and indirect constraints is very effective
in the search for additional neutral Higgs bosons at low Higgs masses and/or
large $\tan \beta$. In fact, we can see that the recent CMS constraints, which discriminate different production mechanisms, reduce the area allowed by the previous ATLAS searches strongly. At present, the second neutral Higgs in a generic MSSM must be heavier than 250~GeV, and such low values for the Higgs mass are possible only for $\tan \beta \simeq 16$. In fact, lower values of $\tan \beta$ require a somewhat heavier neutral Higgs, $m_{H_2}\gtrsim 300$~GeV, due to the large charged Higgs contribution to BR($B\to X_s \gamma$). Larger values of $\tan \beta$ are strongly constrained by the CMS searches in the $H_i \to \tau \tau$ channel and require much heavier Higgs states. For instance, a value of $\tan \beta =30$ would be only possible for $m_{H_2} \gtrsim 600$~GeV. By comparing with the previous estimate from ATLAS results, the improvement becomes aparent. Thus, these bounds are able to constrain very effectively the allowed parameter space in the $(\tan \beta,M_{H_2})$ plane for a generic MSSM, even in the presence of CP violation.  

\begin{figure}
\begin{centering}
\includegraphics[scale=0.8]{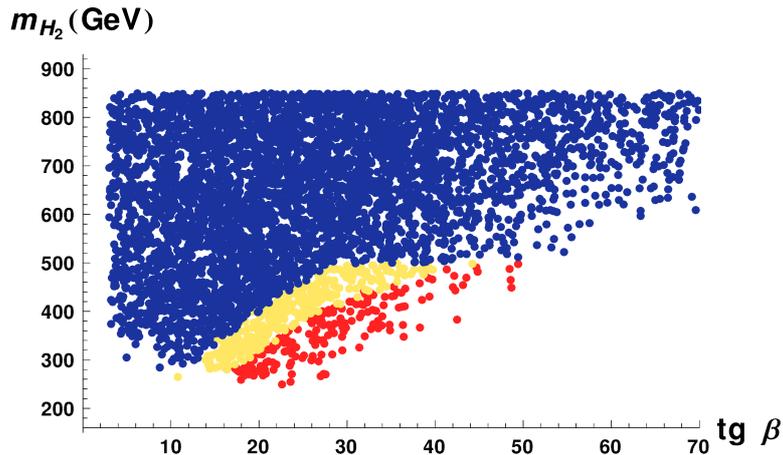}
\par\end{centering}
\caption{\label{fig:futATLASmass}
Allowed Higgs masses in the plane
$(\tan\beta,\: M_{H_2})$ for present and improved ATLAS constraints. Red (dark grey) points satisfy present ATLAS $H_i\to \tau \tau$ bounds
whereas yellow (light grey) and blue (black) points show the effect of improving these bounds on the $\tau \tau$ production cross section by a factor of 5 or 10, respectively. It is clear, that an extension of the analysis up to masses of 1~TeV would be very welcome.}
\end{figure}

Still, as we said in the previous section, it is reasonable to expect ATLAS bounds to improve significantly when the stored data are analyzed. In Figure \ref{fig:futATLASmass}, we present the effect on the allowed values of $(\tan\beta,\:
M_{H_{2}})$ that an improvement of the ATLAS bound  on the $\tau\tau$ production cross section by a factor of 5 or 10  would have. The different colours 
correspond to applying the present
ATLAS bound on $\sigma_{H_i} \times{\rm BR}(H_i\to \tau \tau)$, red
circles, or assuming an improvement of this bound by a factor of
five, yellow circles, or ten, blue circles. These results can also
be applied to the heaviest neutral Higgs, $H_3$, which, in this limit,
is nearly degenerate to $H_2$. We can see here that the present ATLAS $\tau
\tau$ bound is very restrictive for large values of $\tan \beta$,
although the bound is relaxed for heavier Higgs masses and for
$m_{H_2} \gtrsim 400$~GeV, $\tan \beta \simeq 50$ is still
allowed, while there is no constraint for Higgs masses above 500 GeV. 
Improving the ATLAS bound by a factor of 5 or 10 reduces
strongly the allowed parameter space. For instance, an improvement by
a factor of 10 would restrict $\tan \beta <30$, for $m_{H_2} \leq
500$~GeV. Needless to say, an extension of the analysis up to masses of 1 TeV, at least, is not only welcome but absolutely necessary. 

 On the other hand, at low $\tan \beta$, the constraints from
$B\to X_s \gamma$ eliminate light Higgs masses due to the smallness of
the stop-chargino contribution which can not compensate the large
charged Higgs contribution. The combination of $\tau \tau$ and $B\to
X_s \gamma$ constraints implies that $H_{2,3}$ masses below 250~GeV are already ruled out. An improvement of the $\tau \tau$ bound by a factor of 10, which
could be possible with the analysis of the stored LHC data
\cite{privateFiorini}, would eliminate the possibility of $m_{H_2}
\leq 300$~GeV for all $\tan \beta$ values. Thus, we can see that the 
combination of both constraints is very important in the searches for 
additional Higgs states at LHC.

\section{Second Higgs at $m_{H_{2}}=136.5$ GeV}
\label{sec:second}
In his recent study on Higgs resonance properties using the diphoton
channel \cite{CMS-PAS-HIG-13-016}, the CMS collaboration analyzed an
integrated luminosity of 5.1 (19.6) fb$^{-1}$ at a centre of mass
energy of 7 (8) TeV. This analysis searched for a second Higgs-like
state, aside from the signal at 125--126~GeV previously reported and
widely interpreted as the SM Higgs boson, in the range $110 <m_H<
150$~GeV. The result of this analysis reveals a clear excess at
$m_H=136.5$~GeV with a local significance of 2.73 $\sigma $ combining
the data from gluon fusion and vector-boson associated production
(each of which shows the excess individually). Even though there is no
other channel (as $H\to W W^*$, $H\to Z Z^*$, $H\rightarrow\tau\tau$
\dots) backing this result, the statistical analysis has proven to be
incapable of eliminating this particular excess. However, as we have shown in 
this paper, the combination of $\tau\tau$ and BR($B\to X_s \gamma$) constraints 
restricts any additional neutral Higgs in the MSSM to be above 250~GeV. 
Nevertheless, we will show here that it is not possible to accommodate two 
peaks of sizeable strength in the $\gamma\gamma$ channel in an MSSM model, 
even disregarding the $\tau\tau$ and BR($B\to X_s \gamma$) constraints.

In section \ref{sec:analysis}, we have seen that it is difficult to reproduce
the observed $O(1)$ signal strength of the diphoton signal for $\tan
\beta \geq 1$. This is due to the fact that the amplitude for the
process $H_i \to \gamma\gamma$ is basically set by the SM particles
running in the loop, mainly $W$-boson and top quark, while heavy SUSY
particles are typically subdominant. The only way to increase the
diphoton amplitude would be to use large values of $\tan \beta$, when
the down-type couplings, both for fermions and scalars, are enhanced
with respect to their SM value. The stau contributions, which have
been advocated in the literature as a possible solution to this
problem, are only effective for very large $\tan \beta$ values and
light stau masses.  However, although in this case the diphoton
amplitude is increased, the tree-level decays to bottom and tau-lepton
also increase, so that the diphoton branching ratio typically decreases. This
effect can not be compensated by an enhancement of the Higgs
production cross section which would also modify the successful
predictions in other Higgs decay channels. Then, the only possibility
to reproduce the first peak at $m_{H_{1}}=126$~GeV is to reduce the
down-type and pseudoscalar component of $H_1$, and to constrain the
value of $U_{12}$, i.e. the up-type Higgs component of $H_1$, to be
close to unity, as shown in Figure \ref{fig:component-photons}. In
this way, it is possible to reproduce the observed signal strength for
$H_1$, but this implies that the other two neutral Higgs states in the
MSSM have necessarily large down-type and pseudoscalar components.

Then, using this solution to reproduce the first peak in $\gamma\gamma$ 
at $\sim 126$~GeV, it is clear 
that we can not repeat the same strategy to have a
second peak of an intensity similar to the SM one at $m_{H_{2}}\simeq
136$~GeV. This second Higgs state necessarily has a small up-type component, 
which will go as $U_{22}\sim 1 /\tan\beta$, and then $(U_{21}^2 + U_{23}^2)
\sim 1 - 1/\tan^2\beta$. Moreover, $\Gamma (H_2 \to \gamma\gamma)$ has
to be compared with the SM Higgs cross section for $m_H = 126$~GeV and
the $W$-boson contribution to the decay width, dominant for the SM Higgs, 
would be much smaller for $H_2$ with these mixings, as $S_{H_{2}^{0},W}^{\gamma}
\simeq-8.3\,\left(\mathcal{U}_{22}+
\mathcal{U}_{21}/\tan\beta\right)$. So, the W-boson contribution to the 
$H_2$ decay width is
suppressed by a factor $\tan \beta$ and this reduction
of the amplitude can not be compensated by an increase in the contributions 
from down-type
fermions or sfermions to the diphoton triangle with large $\tan
\beta$. For instance, we could think that the $b$-quark contribution to the scalar amplitude, given by 
$S_{H_{2}^{0},b}^{\gamma}\sim \left(-0.025+i\,0.034\right) ~\tan\beta~ (\mathcal{U}_{11} + \tan \beta~ \mbox{Im}\left\{\kappa_{d}\right\} {\cal U}_{13}) $, could compensate the W-boson contribution. However, for typical values $\kappa_d\simeq 0.05$, this would require values of $\tan \beta\geq 80$, while, on the other side, the $H_2 \to b \bar b$ tree-level decay width would also increase with $\tan^2 \beta$ so that the diphoton branching ratio would be decreased. The same reasoning is valid for the case of light staus, which can not contribute significantly to the diphoton scalar amplitude as shown in Figure \ref{figure5}. 

In summary, reproducing two SM-size peaks in the diphoton spectrum is not possible in a generic MSSM setup, even before considering the additional constraints from the $\tau\tau$ and BR($B\to X_s \gamma$) searches. Adding then the present $\tau\tau$ constraints reinforces this conclusion and we can completely discard an MSSM explanation of this second peak in the $\gamma\gamma$ spectrum.

\section{Conclusions}
\label{sec:conclusions}
In this work we have used the latest LHC results on the two photon 
signal strength and the $\tau\tau$ production cross sections, together with 
the indirect low-energy constraints on BR$(B\to X_s \gamma)$, to restrict 
the allowed parameter space of the Higgs sector in a generic MSSM. 

Our study starts with the $\gamma \gamma$ signal observed at LHC at $m_H\simeq 126$~GeV. The experimental results show a signal slightly larger or of the order of the SM expectations, and this is a strong constraint on models with extended Higgs sectors. For large $\tan \beta$ values, when the partial width $\Gamma(H_1 \to \gamma \gamma)$ can be increased, the branching ratio BR$(H_1 \to \gamma \gamma)$ tends to be smaller than the SM if the down-type or pseudoscalar components of $H_2$ are sizeable. Requiring $\sigma(pp\to H_1) \times \mbox{BR}(H_1 \to \gamma \gamma)$ to be of the order of the SM severely restricts the possible mixings in the Higgs sector, so that the down-type or pseudoscalar components of $H_1$ are required to be $\lesssim 1/ \tan \beta$.

Next, we have analyzed the $\tau\tau$ production cross sections for the three Higgs eigenstates. We have shown the present constraints on a generic MSSM coming from $\sigma (pp \to H_i \to \tau \tau)$, including for the first time to the best of our knowledge, the new CMS constraints of neutral MSSM Higgs bosons up to 1~TeV which discriminate different Higgs production mechanisms. As it became apparent in our analysis, the combination of the recent CMS $\tau \tau$ searches and indirect constraints is an excellent weapon in any strategy to search for additional Higgs states at LHC. In this respect, both an update and an extension up to 1 TeV of the present ATLAS analysis is mandatory.
If the theory that is hiding so effectively behind the SM is in fact the MSSM, the $\tau\tau$ searches are the ideal tool to bail it out.

\section*{Acknowledgments}
The authors are grateful to William Bardeen, Luca Fiorini, and Arcadi Santamaria
for useful discussions. We acknowledge support from the MEC and FEDER (EC)
Grants FPA2011-23596 and the Generalitat Valenciana under grant  PROMETEOII/2013/017. G.B. acknowledges partial support from the European Union FP7 ITN INVISIBLES (Marie Curie Actions, PITN-GA-2011-289442). C.B. thanks {\it Ministerio de Educaci\'on, Cultura y Deporte} for finantial support through an FPU-grant AP2010-3316. The work of M.L.L.I is funded through an FPI-grant   BES-2012-053798  from {\it Ministerio de Economia y Competitividad}.

\end{document}